\documentclass[12pt]{article}
\usepackage{float}
\usepackage{eurosym}
\usepackage{graphicx}
\usepackage[utf8]{inputenc}
\usepackage[T1]{fontenc}
\usepackage{indentfirst}
\usepackage[margin=0.6in,nomarginpar]{geometry}
\usepackage[final]{hyperref}
\usepackage{amsmath}
\usepackage{hyperref}
\usepackage{cite}
\usepackage{subcaption}
\usepackage{caption}
\usepackage{amssymb}
\usepackage{multirow}

\usepackage[table]{xcolor}
\usepackage{slashed}
\usepackage{orcidlink}

\hypersetup{
colorlinks=true,
linkcolor=blue,
citecolor=blue,
filecolor=magenta,
urlcolor=blue
}

\begin{document}

\title{Vacuum Pair Creation in Spin Noncommutative of Coordinates: Volkov Background and Constant Electric Field}
\author{B. Hamil \orcidlink{0000-0002-7043-6104} \thanks{%
hamilbilel@gmail.com/bilel.hamil@umc.edu.dz (Corresponding author)} \\
Laboratoire de Physique Math\'{e}matique et Physique Subatomique, LPMPS,\\
Facult\'{e} des Sciences Exactes, Universit\'{e} Constantine 1, Constantine,
Algeria \and B. C. L\"{u}tf\"{u}o\u{g}lu \orcidlink{0000-0001-6467-5005} \thanks{%
bekir.lutfuoglu@uhk.cz } \\
Department of Physics, Faculty of Science, University of Hradec Kralove, \\
Rokitanskeho 62/26, Hradec Kralove, 500 03, Czech Republic.}
\date{}
\maketitle

\begin{abstract}
We investigate the phenomenon of vacuum pair creation for Dirac fermions subjected to a Volkov plane wave and a constant electric field within the framework of spin noncommutativity of coordinates. Employing the Schwinger proper-time formalism, we derive the effective action and obtain closed-form expressions for the pair creation probability. Our analysis reveals that, in the presence of a Volkov plane wave background, the pair production probability remains zero—even with spin noncommutativity. In contrast, for a constant electric field in $(1+1)$-dimensional spacetime, the spin-induced noncommutative deformation significantly enhances the pair creation probability. Remarkably, we identify a critical value of the deformation parameter, $\ell = \frac{m}{eE}$, at which the pair creation probability diverges, indicating a potential vacuum instability or a breakdown of the perturbative regime. These findings underscore the nontrivial role of spin noncommutativity in nonperturbative quantum electrodynamics and offer novel insights for future studies in strong-field physics.
\end{abstract}

\begin{description}
\item[Keywords:] Pair creation; Dirac equation; Spin noncommutativity of coordinates;
Volkov plane wave; Schwinger formalism
\end{description}

\section{Introduction}

Dirac's equation~\cite{Dirac}, which elucidates the behavior of relativistic particles such as electrons, unexpectedly predicted the existence of negative energy states. To address this, Dirac proposed the hole theory, wherein an electron-positron pair is formed when an electron is promoted from the negative energy sea (the Dirac sea) to the positive energy continuum. In 1951, Schwinger~\cite{Schwinger}—and even earlier, Heisenberg and Euler~\cite{Euler}, though from a different perspective—demonstrated that an extremely intense external electric field can spontaneously create particle-antiparticle pairs out of the vacuum. This process, now known as the Schwinger effect, yields a pair production rate per unit volume and unit time given by
\begin{equation}
N=\frac{e^{2}E^{2}}{\left( 2\pi \right) ^{3}}e^{-\frac{\pi m^{2}}{eE}}.
\end{equation}
In this expression, the exponential factor underscores the non-perturbative nature of the phenomenon, becoming significant only when the electric field exceeds a critical threshold. Within the framework of quantum electrodynamics (QED), a magnetic field alone cannot induce pair production, but it can modulate the pair creation rate when combined with an electric field, as shown in~\cite{Karabali1, Karabali2}. Further studies have revealed that a strong magnetic field may also stabilize the vacuum by influencing quantum correlations between created pairs~\cite{Bhattacharya}. Although originally derived in the context of QED~\cite{bilel2, gold2021, Liu}, the Schwinger effect can be extended to other field theories coupled to external fields—for instance, scalar field theories with scalar or vector couplings, and in quantum chromodynamics~\cite{Biro, Kajantie, Kharzeev, McLerran, Venugopalan, Kim2009, Aguilar2023}.

To date, the Schwinger mechanism has not been directly observed, as it requires extremely strong electric fields on the order of \(10^{18}~\text{V/m}\). However, advancements in high-intensity laser technology~\cite{Sarri, Turcu, Ibraam, Kumar2023, Schmitt2023} provide hope for experimental realization. In the meantime, theoretical efforts continue via numerical simulations~\cite{Gies, Dunne, Mocken, Orthaber, Jiang, Aleksandrov} and various analytical techniques, including Hamiltonian diagonalization~\cite{Grib}, path integral methods~\cite{Chitre, Nayak}, worldline instanton techniques~\cite{Schubert, Dunne2}, semiclassical WKB approximations~\cite{Biswas, Shaw, Kim}, and "in"–"out" state formalisms~\cite{Nikishov}. Additionally, pair creation has been explored in different spacetime geometries~\cite{Frob, Kobayashi, Stahl, bilel4, Bavarsad, bilel1, Kim2022}, including those with noncommutative structures~\cite{Grosse, Jabbari, bilel3, bilel5, Mehdaoui, Moures}.

The idea that position operators may not commute, which emerged prominently in string theory~\cite{Connes} and quantum gravity~\cite{Doplicher}, was first formalized by Snyder to address divergences in field theory~\cite{Snyder1, Snyder2}:
\begin{equation}
\left[ \hat{X}_{\mu },\hat{X}_{\nu }\right] =2i\ell ^{2}L _{\mu \nu }, 
\label{sny1}
\end{equation}%
where $L_{\mu\nu}$ are the generators of the Lorentz group, and $\ell$ is a small deformation parameter.

In 2009, Falomir et al.~\cite{Falomir} introduced a three-dimensional noncommutative quantum mechanical system characterized by a deformed algebra mixing spatial and spin operators:
\begin{eqnarray}
    \left[ \hat{X}_{i},\hat{X}_{j}\right] =i \ell ^{2} \varepsilon_{ijk} \hat{S}_{k}, \quad 
    \left[ \hat{S}_{i },\hat{S}_{j }\right] =i \varepsilon_{ijk} \hat{S}_{k}, \quad  \left[ \hat{P}_{i},\hat{P}_{j}\right] =0, \quad
     \left[ \hat{X}_{i},\hat{P}_{j}\right] =i  \delta_{ij}, \quad \left[ \hat{X}_{i},\hat{S}_{j}\right] =i  \ell  \varepsilon_{ijk} \hat{S}_{k}.
\end{eqnarray}
Here, \(\hat{S}_k = \frac{\sigma_k}{2}\) is the spin operator written in terms of the Pauli matrices, and \(\ell\) denotes a deformation parameter with dimensions of length. 

A year later, Gomes et al.~\cite{Gomes} used a similar algebraic framework to generalize the Pauli and Dirac equations. Since then, this spin-dependent noncommutative algebra has been applied to a variety of physical systems, including the Aharonov–Bohm effect, quantum harmonic oscillators, Dirac and DKP oscillators, the inverse square potential, and hydrogen-like atoms~\cite{Das, Falomir2012, Vasyuta2013, Vasyuta2014, Vasyuta2017, Ham2017, Mou2023}. Soon after, in 2016, Vasyuta et al.~\cite{Vasyuta2016} extended this framework by replacing the Pauli matrices with Dirac gamma matrices. 

Despite these advances, the implications of spin-induced noncommutativity for quantum vacuum phenomena—particularly in strong-field quantum electrodynamics—remain largely unexplored. Motivated by this gap, in this work we investigate the Schwinger pair production process within the context of spin noncommutative geometry. Our goal is to understand how spin-dependent deformations of spacetime affect nonperturbative vacuum instabilities, and whether they introduce new mechanisms or amplification channels for pair creation under external fields.

The manuscript is organized as follows. In Section~\ref{sec2}, we outline the formal structure of spin noncommutativity and the corresponding modifications to field operators. Section~\ref{sec3} presents the effective action derived via the Schwinger proper-time formalism and establishes the modified Hamiltonian. In Section~\ref{sec4}, we apply the framework to a Volkov plane wave background and demonstrate the absence of pair production. Section~\ref{sec5} focuses on the case of a constant electric field in $(1+1)$ dimensions, where we compute the pair creation rate analytically and analyze its dependence on the deformation parameter. Finally, Section~\ref{sec6} summarizes our findings and discusses potential directions for future research.


\newpage

\section{Spin noncommutativity of coordinates} \label{sec2}

According to a central assumption in space-time quantization theory, certain points in spacetime do not possess precise conceptual meaning~\cite{Snyder1, Snyder2, 16, 17}. This idea is reflected in models where the commutators of position operators do not vanish. In particular, in quantized spacetime models at small distances, such noncommutativity is often associated with the spin degrees of freedom of quantum fields.

To the best of our knowledge, scalar particles do not exhibit spacetime noncommutativity. However, Dirac fermions can experience such effects. In this work, following the framework proposed by Vasyuta et al.~\cite{Vasyuta2016}, we adopt the following modified commutation relations for a Dirac field:
\begin{equation}
[\hat{X}_\mu, \hat{X}_\nu] = 2i \ell^2 \sigma_{\mu \nu},
\label{es}
\end{equation}
where \(\sigma_{\mu \nu}\) are defined as the commutators of Dirac gamma matrices:
\begin{equation}
\sigma_{\mu \nu} = \frac{1}{2i} [\gamma_\mu, \gamma_\nu].
\end{equation}

The modified position operator \(\hat{X}_\nu\) is constructed as a deformation of the usual Heisenberg operator \(\hat{x}_\nu\), given by:
\begin{equation}
\hat{X}_\nu = \hat{x}_\nu + \ell \gamma_\nu.
\label{rep}
\end{equation}

In this spin-noncommutative geometry, the algebra is closed with the following commutation relations:
\begin{equation}
[\hat{X}_\mu, \hat{p}_\nu] = i g_{\mu \nu}, \quad
[\hat{X}_\mu, \gamma_\nu] = 2i \ell \sigma_{\mu \nu}, \quad
[\hat{p}_\mu, \hat{p}_\nu] = 0, \quad
[\hat{p}_\mu, \hat{\sigma}_{\alpha \beta}] = 0,
\end{equation}
where the Minkowski metric is \(g_{\mu \nu} = \mathrm{diag}(-1, 1, 1, 1)\).

Given a smooth function \(f(\hat{X})\), it can be Taylor-expanded around \(\hat{x}_\mu\) as:
\begin{align}
f(\hat{X}) &= f(\hat{x} + \ell \gamma) = f(\hat{x}) + \ell \gamma^\nu \partial_\nu f(\hat{x}) 
+ \frac{\ell^2}{2!} \langle \gamma^\mu \gamma^\nu \rangle \partial_\mu \partial_\nu f(\hat{x}) + \frac{\ell^3}{3!} \langle \gamma^\mu \gamma^\nu \gamma^\rho \rangle \partial_\mu \partial_\nu \partial_\rho f(\hat{x}) + \cdots,
\label{dev}
\end{align}
where the angular brackets denote symmetrization over all permutations of indices:
\begin{equation}
\langle \gamma_{\mu_1} \gamma_{\mu_2} \cdots \gamma_{\mu_n} \rangle = \frac{1}{n!} \sum_{\text{perm}} \gamma_{\mu_1} \gamma_{\mu_2} \cdots \gamma_{\mu_n}.
\end{equation}

In the non-relativistic limit, the gamma matrices \(\gamma_\mu\) reduce to the Pauli matrices \(\sigma_i\), and the deformed algebra simplifies accordingly. In this case, the noncommutative position operators become matrix-valued and are expressed as:
\begin{equation}
\hat{X}_i = \hat{x}_i + \ell s_i, \quad \text{with} \quad s_i = \frac{\sigma_i}{2}, \quad i = 1, 2, 3.
\end{equation}

As a consequence, the Schrödinger equation acquires a spin-dependent correction and takes the form:
\begin{equation}
\left[ \frac{p^2}{2m} + V(\hat{x}_i + \ell s_i) \right] \psi = E \psi,
\end{equation}
where \(\psi\) is the Pauli spinor.

\section{Effective action in a space with spin noncommutativity of coordinates} \label{sec3}

We now consider the dynamics of a massive Dirac particle interacting with an external electromagnetic field. The action is given by
\begin{equation}
\mathcal{S} = \int d^{4}x \left[ \bar{\psi} \left( \hat{\slashed{p}} - e\slashed{A} \right) \psi - m\bar{\psi} \psi \right],
\label{act}
\end{equation}
where \(\psi\) is the Dirac spinor, \(\bar{\psi}\) its Dirac adjoint, and \(\slashed{a} = \gamma^{\mu} a_{\mu}\). 

To incorporate spin noncommutativity, we treat the gauge field $A_{\mu
}\left( \hat{X}\right) $ as a matrix-valued function of the operator $\hat{X}%
^{\mu }=\hat{x}^{\mu }+\ell \gamma ^{\mu }$. Expanding $A_{\mu }\left( \hat{X%
}\right) $ around the commutative coordinate $\hat{x}^{\mu }$, we obtain

\begin{equation}
A_{\mu }\left( \hat{X}\right) =A_{\mu }\left( \hat{x}\right) +\mathcal{B}_{\mu }\left( \hat{x}\right), \label{fie}
\end{equation}
where%
\begin{equation}
\mathcal{B}_{\mu }\left( \hat{x}\right) =\ell \gamma _{\nu }\partial ^{\nu}A_{\mu }\left( \hat{x}\right) +\frac{\ell ^{2}}{2!}\left\langle \gamma_{\alpha }\gamma _{\beta }\right\rangle \partial ^{\alpha }\partial ^{\beta}A_{\mu }\left( \hat{x}\right) +\frac{\ell ^{3}}{3!}\left\langle \gamma_{\nu }\gamma _{\alpha }\gamma _{\beta }\right\rangle \partial ^{\nu}\partial ^{\alpha }\partial ^{\beta }A_{\mu }\left( \hat{x}\right) +\cdots.
\end{equation}

Substituting Eq.~\eqref{fie} into Eq.~\eqref{act}, we obtain the modified action in the presence of spin noncommutativity:
\begin{equation}
\mathcal{S} = \int d^4x \left[ \bar{\psi} \left( \hat{\slashed{p}} - e\slashed{A} - e\gamma^{\mu} \mathcal{B}_\mu(\hat{x}) \right) \psi - m\bar{\psi} \psi \right].
\end{equation}

In the limit \(\ell \rightarrow 0\), the noncommutative correction vanishes and the standard commutative Dirac action is recovered, as expected.

To evaluate the vacuum-to-vacuum transition amplitude, we perform the path integral over the Dirac spinor fields:
\begin{align}
\mathcal{A}(\text{vac} \rightarrow \text{vac}) 
&= \mathcal{N} \int D\psi D\bar{\psi} \, 
\exp \left\{ i \int d^4x \left[ \bar{\psi} \left( \hat{\slashed{p}} - e\slashed{A} - e\gamma^{\mu} \mathcal{B}_\mu(\hat{x}) \right) \psi - m\bar{\psi} \psi \right] \right\} \notag \\
&= \exp \left[ -\operatorname{tr} \ln \left( \frac{\hat{\slashed{p}} - m + i\varepsilon}{\hat{\slashed{p}} - e\slashed{A} - e\gamma^{\mu} \mathcal{B}_\mu(\hat{x}) - m + i\varepsilon} \right) \right],
\label{vac}
\end{align}
where \(\mathcal{N}\) is a normalization constant that can be determined by comparison with the free (non-interacting) case. Here, the trace operation \(\operatorname{tr} = \operatorname{tr}_x \operatorname{tr}_\gamma\) includes both a trace over the configuration space (\(\operatorname{tr}_x\)) and a trace over spinor indices (\(\operatorname{tr}_\gamma\)).

The Dirac equation is invariant under charge conjugation, satisfying the identity \(C^{-1} \gamma_\mu C = -\gamma_\mu^T\), which allows us to recast the amplitude as:
\begin{equation}
\mathcal{A}(\text{vac} \rightarrow \text{vac}) = \exp \left[ -\frac{1}{2} \operatorname{tr} \ln \left( \frac{\hat{p}^2 - m^2 - i\varepsilon}{\left( \hat{\slashed{p}} - e\slashed{A} - e\gamma^{\mu} \mathcal{B}_\mu(\hat{x}) \right)^2 - m^2 - i\varepsilon} \right) \right].
\end{equation}

Using the operator identity
\begin{equation}
\ln \left( \frac{f}{g} \right) = \int_0^{\infty} \frac{ds}{s} \left( e^{isg} - e^{isf} \right),
\end{equation}
we rewrite the amplitude in exponential form:
\begin{equation}
\mathcal{A}(\text{vac} \rightarrow \text{vac}) = e^{i S_{\text{eff}}},
\end{equation}
where the one-loop effective action is given by
\begin{equation}
S_{\text{eff}} = \frac{i}{2} \operatorname{tr} \int_0^{\infty} \frac{ds}{s} \left\{ G(x_b, x_a, s) - e^{is(\hat{p}^2 - m^2)} \right\}_{x_b = x_a}.
\label{ay}
\end{equation}

Here, \(G(x_b, x_a, s)\) denotes the proper-time propagator:
\begin{equation}
G(x_b, x_a, s) = \langle x_a | e^{is \mathcal{H}} | x_b \rangle,
\end{equation}
where the Hamiltonian operator \(\mathcal{H}\) reads:
\begin{equation}
\mathcal{H} = \left( \hat{\slashed{p}} - e\slashed{A} - e\gamma^{\mu} \mathcal{B}_\mu(\hat{x}) \right)^2 - m^2.
\label{Hm1}
\end{equation}

\section{Volkov plane wave solution} \label{sec4}

In this section, we investigate the possibility of vacuum pair creation in the presence of a Volkov-type plane wave background within the framework of spin noncommutative coordinates. The Volkov solution provides the exact wavefunction for a charged particle in a classical plane electromagnetic wave and serves as a crucial test case in strong-field QED. Our objective is to derive the modified propagator under spin noncommutativity and to determine whether it alters the well-established result that Volkov fields do not induce particle-antiparticle pair creation.

In the literature, the Volkov plane wave field \(A_\mu\) is characterized by the following properties:
\begin{itemize}
    \item \(A_\mu\) is a function of the scalar product \(kx\), where \(k^\mu\) is the four-wavevector of the plane wave that satisfies \(k^2 = k_\mu k^\mu = 0\).
    \item \(A_\mu\) satisfies the transversality (Lorentz gauge) condition: \(\partial^\mu A_\mu = k^\mu \dot{A}_\mu = 0\), where $\dot{A}_\mu = \frac{\partial A_{\mu }}{\partial kx}$.
\end{itemize}

In the presence of spin noncommutativity, the Volkov plane wave field becomes
\begin{equation}
A_\mu(k\hat{X}) = A_\mu(k\hat{x} + \ell \slashed{k}).
\end{equation}

By employing the noncommutative position operators  $\hat{X}$ that satisfy
the commutation relation (\ref{rep}), the plane-wave field $%
A_{\mu }\left( k\hat{X}\right) $ can be written as a Taylor expansion around
the commutative coordinate:
\begin{equation}
A_{\mu }\left( k\hat{X}\right) =A_{\mu }\left( k\hat{x}\right) +\ell \gamma
_{\nu }k^{\nu }\frac{\partial A_{\mu }}{\partial A_{\mu }\left( kx\hat{x}\right) }+\frac{%
\ell ^{2}}{2!}%
\left\langle \gamma _{\alpha }\gamma _{\beta }\right\rangle
 k^{\alpha 
}k^{\beta }\frac{\partial ^{2}A_{\mu }}{\partial A_{\mu }\left( kx\hat{x}\right)
 ^{2}}++\frac{\ell ^{3}}{3!}\left\langle 
\gamma _{\nu }\gamma _{\alpha }\gamma
 _{\beta }\right\rangle k^{\nu 
}k^{\alpha }k^{\beta }\frac{\partial
^{3}A_{\mu }}{\partial A_{\mu }\left( kx\hat{x}\right) ^{3}}+...
\end{equation}%
Equivalently,%
\begin{equation}
A_{\mu }\left( k\hat{X}\right) =A_{\mu }\left( k\hat{x}\right) +\ell %
\slashed{k}\frac{\partial A_{\mu }}{\partial A_{\mu }\left( kx\hat{x}\right) }+\frac{\ell
 ^{2}}{2!}\left\langle %
\slashed{k}\slashed{k}\right\rangle \frac{\partial
^{2}A_{\mu }}{\partial A_{\mu }\left( kx\hat{x}\right) ^{2}}+\frac{%
\ell ^{3}}{3!}%
\left\langle \slashed{k}\slashed{k}\slashed{k}\right\rangle \frac{\partial
^{3}A_{\mu }}{\partial 
A_{\mu }\left( kx\hat{x}\right) ^{3}}+...
\end{equation}

Using the Clifford algebra $\left\{ \gamma ^{\mu },\gamma ^{\nu }\right\} =2g^{\mu \nu
}$ together with the light-like condition for the Volkov wave vector $%
k^{2}=k_{\mu }k^{\mu }=0$ , all higher-order contractions satisfy%
\begin{equation}
\left\{ 
\begin{array}{c}
\left\langle \gamma _{\alpha }\gamma _{\beta }\right\rangle k^{\alpha
}k^{\beta }=\left\langle \slashed{k}\slashed{k}\right\rangle =\frac{1}{2}%
\left( 2g_{\alpha \beta }\right) k^{\alpha }k^{\beta }=k^{2}=0 \\ 
\left\langle \gamma _{\nu }\gamma _{\alpha }\gamma _{\beta }\right\rangle
k^{\nu }k^{\alpha }k^{\beta }=\left\langle \slashed{k}\slashed{k}\slashed{k}%
\right\rangle =k^{2}\slashed{k} \\ 
\text{etc}%
\end{array}%
\right. .
\end{equation}%
Therefore, the full Taylor series collapses to a single non-vanishing
contribution:

\begin{equation}
A_{\mu }\left( k\hat{X}\right) =A_{\mu }\left( k\hat{x}\right) +\ell %
\slashed{k}\dot{A}_{\mu }.  \label{ex}
\end{equation}%
Substituting Eq. (\ref{ex}) into the Hamiltonian $\mathcal{H}$, and using the identity \(\slashed{A}\slashed{B} + \slashed{B}\slashed{A} = 2 A \cdot B\), we obtain:

\begin{align}
\mathcal{H} 
&= \left( \hat{\slashed{p}} - e\slashed{A} + \ell \slashed{k} \dot{\slashed{A}} \right)^2 - m^2 \notag \\
&= (\hat{p} - eA)^2 - i e \slashed{k} \dot{\slashed{A}} 
+ 2e\ell \dot{\slashed{A}} (k \cdot \hat{p}) - 2e\ell \slashed{k} (\dot{A} \cdot \hat{p}) 
+ 2e\ell \slashed{k} \dot{\slashed{A}} \hat{\slashed{p}} 
+ 2e^2 \ell \slashed{k} A \cdot \dot{A} - 2e^2 \ell \slashed{k} \dot{\slashed{A}} \slashed{A} - m^2.
\end{align}

Thus, the propagator for a Dirac particle in this background is given by
\begin{align}
G(x_b, x_a, s) = \langle x_a | \exp \Bigl\{ is \Bigl[ &(\hat{p} - eA)^2 - i e \slashed{k} \dot{\slashed{A}} + 2e\ell \dot{\slashed{A}} (k \cdot \hat{p}) - 2e\ell \slashed{k} (\dot{A} \cdot \hat{p}) \notag \\
&+ 2e\ell \slashed{k} \dot{\slashed{A}} \hat{\slashed{p}} + 2e^2 \ell \slashed{k} A \cdot \dot{A} - 2e^2 \ell \slashed{k} \dot{\slashed{A}} \slashed{A} - m^2 \Bigr] \Bigr\} | x_b \rangle.
\label{pro}
\end{align}

Before proceeding with the calculations, it is useful to outline our strategy for tackling the problem. Taking advantage of the fact that the plane wave field depends solely on the variable $k\hat{x}$, we introduce a change of variables that treasts this quantity as independent from the spacetime coordinate $x$. This transformation enables us to decouple the free quadridimensional exterior motion from the associated terms, thereby reducing the problem to unidimensional ones that are closely tied to classical motion. For this, we define a new operator $\hat{\phi}=k\hat{x}$, whose action on its eigenstates is given by $\hat{\phi} \left\vert \phi \right\rangle = \phi \left\vert \phi \right\rangle$, and associate $\hat{\phi}$ with its canonically conjugate momentum operator.
\begin{equation}
[\hat{\phi}, \hat{p}_\phi] = i, \quad [\hat{p}_\phi, \hat{x}_\mu] = [\hat{p}_\phi, \hat{p}_\mu] = 0, \quad [\hat{\phi}, \hat{x}_\mu] = [\hat{\phi}, \hat{p}_\mu] = 0.
\end{equation}

To get the free part of the propagator, we introduce the delta functional identity into Eq.~(\ref{pro})
\begin{equation}
\int d\phi_b d\phi_a \delta(\phi_a - k x_a) \delta(\phi_b - \phi_a - k (x_b - x_a)) = \int d\phi_b d\phi_a \delta(\phi_a - k x_a) \langle \phi_b | \phi_a \rangle,
\end{equation}
and perform the shift \(\hat{p}_\mu \rightarrow \hat{p}_\mu + k_\mu \hat{p}_\phi\). Using the relations \(k^2 = k \cdot A = k \cdot \dot{A} = 0\), we then arrive at the following
expression:
\small
\begin{eqnarray}
G &=&\int d\phi _{b}d\phi _{a}\delta \left( \phi _{a}-kx_{a}\right)
\left\langle \phi _{a},x_{a}\right\vert \exp \Biggl\{is\Biggr[\hat{p}%
^{2}-m^{2}+2k\hat{p}\hat{p}_{\phi }\Biggl(1+  \notag \\
&&\frac{1}{\hat{p}_{\phi }}\frac{e^{2}A^{2}-2eA\hat{p}-ie\slashed{k}\dot{{%
\slashed{A}}}+2e\ell \dot{{\slashed{A}}}\left( k\hat{p}\right) -2e\ell %
\slashed{k}\left( \dot{A}\hat{p}\right) +2e\ell \slashed{k}\dot{{\slashed{A}}%
}\hat{\slashed{p}}+2e^{2}\ell \slashed{k}A\dot{A}-2e^{2}\ell \slashed{k}\dot{%
{\slashed{A}}}\slashed{A}}{2k\hat{p}}\Biggl)\Biggr]\Biggl\}\left\vert
x_{b},\phi _{b}\right\rangle .  \label{prs}
\end{eqnarray}%
\normalsize
which is equivalent to 
\begin{equation}
G=\int d\phi _{b}d\phi _{a}\delta \left( \phi _{a}-kx_{a}\right)
\left\langle \phi _{a},x_{a}\right\vert \exp \left\{ is\left[ \hat{p}%
^{2}-m^{2}+2k\hat{p}\hat{p}_{\phi }\left( 1+i\int^{\hat{\phi}}\hat{S}%
_{1}\left( u\right) du\right) \right] \right\} \left\vert x_{b},\phi
_{b}\right\rangle , 
\end{equation}
where
\begin{equation}
\hat{S}_{1}\left( \phi \right) =\frac{e^{2}A^{2}-2eA\hat{p}-ie\slashed{k}%
\dot{{\slashed{A}}}+2e\ell \dot{{\slashed{A}}}\left( k\hat{p}\right) -2e\ell %
\slashed{k}\left( \dot{A}\hat{p}\right) +2e\ell \slashed{k}\dot{{\slashed{A}}%
}\hat{\slashed{p}}+2e^{2}\ell \slashed{k}A\dot{A}-2e^{2}\ell \slashed{k}\dot{%
{\slashed{A}}}\slashed{A}}{2k\hat{p}}.
\end{equation}
Taking into account the identity
\begin{equation}
e^{-i\int^{\hat{\phi}}\hat{S}_{1}\left( u\right) du}\hat{p}_{\phi }e^{i\int^{%
\hat{\phi}}\hat{S}_{1}\left( u\right) du}=\hat{S}_{1}\left( \phi \right) +%
\hat{p}_{\phi }, 
\end{equation}%
in Eq. (\ref{prs}), we get
\small
\begin{equation}
G=\int d\phi _{b}d\phi _{a}\delta \left( \phi _{a}-kx_{a}\right) \delta
\left( \phi _{b}-\phi _{a}-k\left( x_{b}-x_{a}\right) \right) \left\langle
x_{a}\right\vert e^{-i\int^{\hat{\phi}_{a}}\hat{S}_{1}\left( u\right)
du}e^{is\left[ \hat{p}^{2}-m^{2}+2k\hat{p}\hat{p}_{\phi }\right] }e^{i\int^{%
\hat{\phi}_{b}}\hat{S}_{1}\left( u\right) du}\left\vert x_{b}\right\rangle .
\end{equation}
\normalsize
Then, we perform the shift, $\hat{p}\rightarrow \hat{p}-k\hat{p}_{\phi }$, and utilize the following integral representation of the Dirac delta function
\begin{equation}
\delta \left( \phi _{b}-\phi _{a}-k\left( x_{b}-x_{a}\right) \right) =\int 
\frac{dp_{\phi _{b}}}{2\pi }\exp \left[ ip_{\phi _{b}}\left( \phi _{b}-\phi
_{a}-k\left( x_{b}-x_{a}\right) \right) \right] ,
\end{equation}%
Subsequently, by carrying out the integration over the variables $\phi _{b}$ and $\phi _{a}$, we derive the following expression:
\begin{equation}
G=\left\langle x_{a}\right\vert e^{-i\int^{k\hat{x}_{a}}\hat{S}_{1}\left(
u\right) du}e^{is\left( \hat{p}^{2}-m^{2}\right) }e^{i\int^{kx_{b}}\hat{S}%
_{1}\left( u\right) du}\left\vert x_{b}\right\rangle .
\end{equation}
Here, the elimination of the operator $\hat{x}$ is straightforward. since it transforms into a variable $x_{i}$ when operating on the ket $\left\vert x_{i}\right\rangle $. However, the operator $\hat{p}$ remains. To eliminate it, we introduce the projector $\int \left\vert p\right\rangle \left\langle
p\right\vert dp=1$, and utilize the scalar product $\left\langle x\right. \left\vert p\right\rangle =\frac{e^{ipx}}{\left( 2\pi \right) ^{2}}$, and
we get the kernel in the following formula:
\begin{eqnarray}
G\left( x_{b},x_{a},s\right)  &=&\int \frac{d^{4}p}{\left( 2\pi \right) ^{4}}%
\exp \Bigg[-ip\left( x_{b}-x_{a}\right) +is\left( p^{2}-m^{2}\right) +\frac{e%
}{2kp}\slashed{k}\left[ {\slashed{A}}\left( kx_{b}\right) -{\slashed{A}}%
\left( kx_{a}\right) \right]   \notag \\
&&+ie\ell \left[ {\slashed{A}}\left( kx_{b}\right) -{\slashed{A}}\left(
kx_{a}\right) \right] -\frac{i\ell e}{kp}\slashed{k}\left[ A_{\mu }\left(
kx_{b}\right) -A_{\mu }\left( kx_{a}\right) \right] p^{\mu }  \notag \\
&&+\frac{ie\ell }{kp}\slashed{k}\left[ {\slashed{A}}\left( kx_{b}\right) -{%
\slashed{A}}\left( kx_{a}\right) \right] \slashed{p}+\frac{ie^{2}\ell }{2kp}%
\slashed{k}\left[ A^{2}\left( kx_{b}\right) -A^{2}\left( kx_{a}\right) %
\right]   \notag \\
&&-i\int_{kx_{a}}^{kx_{b}}\frac{2eAp-e^{2}A^{2}+2e^{2}\ell \slashed{k}\dot{{%
\slashed{A}}}\slashed{A}}{2kp}du\Bigg].  \label{prf}
\end{eqnarray}
This expression represents the propagator of a Dirac particle interacting with a plane wave field in a space with spin noncommutativity of coordinates. It is worth noting that the propagator includes an additional correction term that depends on the deformation parameter that arises due to the standard Heisenberg algebra modification. In the absence of the spin noncommutativity deformation, $\ell =0$, Eq.~(\ref{prf}) simplifies to
\begin{eqnarray}
G\left( x_{b},x_{a},s\right) &=&\int \frac{d^{4}p}{\left( 2\pi \right) ^{4}}%
\exp \Bigg[-ip\left( x_{b}-x_{a}\right) +is\left( p^{2}-m^{2}\right) +\frac{e%
}{2kp}\slashed{k}\left[ {\slashed{A}}\left( kx_{b}\right) -{\slashed{A}}%
\left( kx_{a}\right) \right]  \notag \\
&&-i\int_{kx_{a}}^{kx_{b}}\frac{2eAp-e^{2}A^{2}}{2kp}du\Bigg],
\end{eqnarray}
which represents the propagator describing a Dirac particle interaction with a plane wave field. This result agrees with that of \cite{Chetouani}. After determining the propagator, we now calculate the probability of particle-antiparticle pair production occurring in a vacuum. Replacing the propagator from Eq. (\ref{prf}) in the effective action given in Eq. (\ref{ay}), we arrive at
\begin{eqnarray}
S_{eff} &=&\frac{i}{2}tr\int_{0}^{\infty }\frac{ds}{s}\int \frac{d^{4}p}{%
\left( 2\pi \right) ^{4}}\Biggl\{  \notag \\
&&\exp \Biggr[-ip\left( x_{b}-x_{a}\right) +is\left( p^{2}-m^{2}\right)
-i\int_{kx_{a}}^{kx_{b}}\frac{2eAp-e^{2}A^{2}+2e^{2}\ell \slashed{k}\dot{{%
\slashed{A}}}\slashed{A}}{2kp}du  \notag \\
&&+\frac{e}{2kp}\slashed{k}\left[ {\slashed{A}}\left( kx_{b}\right) -{%
\slashed{A}}\left( kx_{a}\right) \right] +ie\ell \left[ {\slashed{A}}\left(
kx_{b}\right) -{\slashed{A}}\left( kx_{a}\right) \right] -\frac{i\ell e}{kp}%
\slashed{k}\left[ A_{\mu }\left( kx_{b}\right) -A_{\mu }\left( kx_{a}\right) %
\right] p^{\mu }  \notag \\
&&+\frac{ie\ell }{kp}\slashed{k}\left[ {\slashed{A}}\left( kx_{b}\right) -{%
\slashed{A}}\left( kx_{a}\right) \right] \slashed{p}+\frac{ie^{2}\ell }{2kp}%
\slashed{k}\left[ A^{2}\left( kx_{b}\right) -A^{2}\left( kx_{a}\right) %
\right] \Biggr]  \notag \\
&&-\exp \left[ -ip\left( x_{b}-x_{a}\right) +is\left( p^{2}-m^{2}\right) %
\right] \Biggl\}_{x_{a}=x_{b}}\text{.}
\end{eqnarray}
Next, we take the trace over the configuration space $tr_{x}=\int d^{4}x\left\langle x\right\vert (.)\left\vert x\right\rangle $. We observe that the exponential's exponent becomes zero, meaning that the effect of the spin noncommutativity of coordinates disappears. Consequently, the effective action vanishes, $S_{eff}=0$, and the probability of pair creation of Dirac particles becomes zero, $\mathcal{P}=0$. This implies that a Volkov plane wave field in the presence of spin noncommutativity of coordinates cannot produce pairs of spinning particles from the vacuum.  This result is consistent with the well-known fact in standard QED that a single plane wave electromagnetic field such as the Volkov solution does not lead to pair production, as discussed in standard quantum field theory textbooks such as Itzykson and Zuber \cite{Itzykson}.

\section{Pair-creation probability within a (1+1) dimensional  constant electric field} \label{sec5}

In this section, we analyze the Schwinger pair production process in a simplified $(1+1)$-dimensional spacetime, where a constant electric field is applied along the $z$-direction. This dimensional reduction permits an exact and transparent computation of the pair creation probability, while still capturing the essential features of spin noncommutativity.

We begin by expressing the vector potential in the presence of spin noncommutative coordinates using Eqs.~(\ref{rep}) and~(\ref{dev}). Since the constant electric field potential is linear in the deformed coordinate $\hat{Z}=\hat{z}+\ell \gamma^{3}$, its noncommutative form is obtained exactly as
\begin{equation}
A_{0}\left( \hat{Z} \right) = -E\hat{Z} = -E\hat{z} + \ell E\gamma^3,
\end{equation}
where $E$ is the strength of the constant electric field which can take positive or negative values. We emphasize that this identity follows directly from the full Taylor expansion of $A_{0}\left( \hat{Z} \right) $ and thus already accounts for all orders in the deformation parameter $\ell$.

The corresponding Hamiltonian becomes
\begin{align}
\mathcal{H} &= \left( \hat{\slashed{p}} - e\slashed{A}\left( \hat{X} \right) \right)^2 - m^2 
= \left( \gamma^0 (p_t + eEz) + \gamma^3 p_z - eE\ell \gamma^0 \gamma^3 \right)^2 - m^2 \notag \\
&= \left( p_t + eEz \right)^2 + i \gamma^0 \gamma^3 eE - p_z^2 + e^2 E^2 \ell^2 - m^2.
\end{align}
This leads to the following expression for the propagator:
\begin{equation}
G\left( x_b, x_a, s \right) = \langle x_a | e^{is \left[ (p_t + eEz)^2 + i\gamma^0\gamma^3 eE - p_z^2 + e^2 E^2 \ell^2 - m^2 \right]} | x_b \rangle.
\end{equation}

To evaluate this propagator, we adopt a path integral formulation by discretizing the exponential as $e^{is\mathcal{H}} = \left( e^{i\varepsilon s\mathcal{H}} \right)^N$ with $\varepsilon = 1/N$. We insert $N-1$ resolutions 
We then employ path integral techniques and we consider $e^{is\mathcal{H}}=\left[ e^{i\varepsilon s\mathcal{H}}\right]^{N}$  with $\varepsilon =1/N$. After that, we insert $N-1$ identities $\int\left\vert x_{j}\right\rangle dx_{j}\left\langle x_{j}\right\vert =1$ between each pair of $e^{i\varepsilon s\mathcal{H}}$ operators. Next, we introduce $N$ momentum integrals $\int \left\vert p_{j}\right\rangle dp_{j}\left\langle p_{j}\right\vert =1$, while taking into account 
\begin{equation}
\left\langle x_{j}\right. \left\vert p_{j}\right\rangle =\frac{d^{2}p_{j}}{%
2\pi }e^{ip_{j}x_{j}},
\end{equation}%
Using the steps outlined above, we express the matrix element of $G\left(x_{b},x_{a},s\right) $  in a continuous form as a path integral over position and momentum variables as follows:
\begin{equation}
G\left( x_{b},x_{a},s\right) =e^{is\left[ i\gamma ^{0}\gamma
^{3}eE+e^{2}E^{2}\ell ^{2}-m^{2}\right] }\mathcal{G}\left(
x_{b},x_{a},s\right) ,
\end{equation}%
where%
\begin{equation}
\mathcal{G}\left( x_{b},x_{a},s\right) =\int dtDz\int Dp_{t}Dp_{z}\exp
\left\{ i\int_{0}^{1}\left[ p_{t}\dot{t}-p_{z}\dot{z}+s\left[ \left(
p_{t}+eEz\right) ^{2}-p_{z}^{2}\right] \right] d\tau \right\} ,
\end{equation}
and $x=\left( t,z\right) $ satisfies the boundary conditions
\begin{equation}
x\left( 0\right) =x_{a}, \quad \text{ and } \quad x\left( 1\right) =x_{b}.
\end{equation}
After expressing $\mathcal{G}\left( x_{b},x_{a},s\right) $ in the continuous path integral, we can perform some of the integrations. First, we integrate over the path $t$ and we find that the momenta $p_{t}$ as constants, i.e., $p_{t}=c^{st}$. Next, we integrate once more over the momentum variable $p_{z}$, and then we  rewrite the propagator as
\begin{equation}
\mathcal{G}\left( x_{b},x_{a},s\right) =\int \frac{dp_{t}}{2\pi }\int Dz\exp
\left\{ ip_{t}(t_{f}-t_{i})+i\int \left[ \frac{\dot{z}^{2}}{4s}%
+se^{2}E^{2}\left( z+\frac{p_{t}}{eE}\right) ^{2}\right] d\tau \right\} .
\end{equation}
This propagator, whose action is quadratic, can be determined directly by
integration or by other methods. In order to avoid tedious calculations, we
use the result given in \cite{Chetouani}, which let us to find
\small
\begin{eqnarray}
\mathcal{G}\left( x_{b},x_{a},s\right) &=&\int \frac{dp_{t}}{2\pi }\sqrt{%
\frac{eE}{2i\pi \sinh \left( 2seE\right) }}\exp \Biggl\{%
ip_{t}(t_{b}-t_{a})+i2eE\left[ \left( z_{a}+\frac{p_{t}}{eE}\right)
^{2}+\left( z_{b}+\frac{p_{t}}{eE}\right) ^{2}\right] \coth \left(
2seE\right)  \notag  \\
&&-ieE\frac{\left( z_{a}+\frac{p_{t}}{eE}\right) \left( z_{b}+\frac{p_{t}}{eE%
}\right) }{\sinh \left( 2seE\right) }\Biggl\}. 
\end{eqnarray}
\normalsize
After determining $\mathcal{K}\left( x_{b},x_{a},s\right) $, we easily obtain the effective action
\begin{eqnarray}
S_{eff} &=&\frac{i}{2}tr\int_{0}^{\infty }\frac{ds}{s}\Biggl\{\int \frac{%
dp_{t}}{2\pi }\sqrt{\frac{eE}{2i\pi \sinh \left( 2seE\right) }}\exp \Biggr[%
-s\gamma ^{0}\gamma ^{3}eE+ise^{2}E^{2}\ell ^{2}-ism^{2}+ ip_{t}(t_{b}-t_{a}) \notag \\
&+&ieE\frac{\left( z+\frac{p_{t}}{eE}\right) ^{2}\left[
\cosh \left( 2seE\right) -1\right] }{\sinh \left( 2seE\right) }\Biggr]-\int 
\frac{dp_{t}}{2\pi }\int \frac{dp_{z}}{2\pi }\exp \left[ is\left(
p_{t}^{2}-p_{z}^{2}-m^{2}\right) \right] \Biggl\} .\,\,\,\,\,  \label{see}
\end{eqnarray}%
Next, we take the integration over $p_{t}$ and $p_{z}$, and we find%
\begin{equation}
S_{eff}=\frac{i}{2}tr\int_{0}^{\infty }\frac{ds}{s}e^{-ism^{2}}\left[ \frac{%
\left\vert eE\right\vert }{4\pi }\frac{e^{-s\gamma ^{0}\gamma
^{3}eE+ise^{2}E^{2}\ell ^{2}}}{\sinh \left( seE\right) }-\frac{1}{4\pi s}%
\right] .  \label{sef}
\end{equation}
We now have to take the traces, first on the Dirac matrix and then on the
configuration space. Then, we use the properties of the Dirac gamma matrices, and we get
\begin{equation}
e^{-s\gamma ^{0}\gamma ^{3}eE}=\cosh \left( seE\right)-\gamma ^{0}\gamma ^{3}\sinh \left( seE\right),
\end{equation}%
Since $tr \, \gamma ^{0}\gamma ^{3}=0$, we can evaluate the trace of the above expression directly. We find
\begin{equation}
tre^{-s\gamma ^{0}\gamma ^{3}eE}=2\cosh \left( seE\right). \label{son1}
\end{equation}%
Substituting Eq. \eqref{son1} in Eq. (\ref{sef}), we obtain%
\begin{equation}
S_{eff}=\frac{i}{4\pi }tr_{x}\int_{0}^{\infty }\frac{ds}{s}e^{-ism^{2}}\left[
\frac{\left\vert eE\right\vert e^{ise^{2}E^{2}\ell ^{2}}}{\tanh \left(
s\left\vert eE\right\vert \right) }-\frac{1}{s}\right] .
\end{equation}%
We now use the identity 
\begin{equation}
 \mathcal{P}=1-e^{i\left( S_{eff}-S_{eff}^{\ast
}\right) }\simeq 2\text{Im}S_{eff}.   
\end{equation}
Here, we need to make the change of variable $s\rightarrow -s$ in the complex conjugate of the effective action $S_{eff}^{\ast}$. Then, we combine all of these manipulations to express the probability 
\begin{equation}
\mathcal{P}=\frac{{\color{black}TL}}{4\pi }\int_{-\infty }^{+\infty }\frac{ds}{s}%
e^{-ism^{2}}\left[ \frac{\left\vert eE\right\vert e^{ise^{2}E^{2}\ell ^{2}}}{%
\tanh \left( s\left\vert eE\right\vert \right) }-\frac{1}{s}\right] .
\end{equation}%
We note that this integral can be evaluated with the help of the residue theorem. The integral has singularities at $s=0$ and simple poles located on the imaginary axis at $s_{n}=-\frac{in\pi }{eE}\left( n\neq 0\right) .$ Here, we close the contour of integration at infinity by a semicircle in the upper half-plane and evaluate the integral by the residue theorem. The result turns out to be:
\begin{equation}
\mathcal{P}=\frac{\left\vert eE\right\vert }{2\pi }\sum_{n=1}\frac{1}{n}e^{-%
\frac{n\pi }{\left\vert eE\right\vert }(m^{2}-e^{2}E^{2}\ell ^{2})}.
\label{proo}
\end{equation}%
This result resembles the well-known result corresponding to particle creation in the (1+1) dimensional space-time, with the change $m^{2}-e^{2}E^{2}\ell ^{2}\rightarrow \mu ^{2}.$ The convergence of the series is obvious. This can also be written in a closed form as:
\begin{equation}
\mathcal{P}=-\frac{\left\vert eE\right\vert }{2\pi }\log \left[ 1-e^{-\pi (%
\frac{m^{2}}{\left\vert eE\right\vert }-\left\vert eE\right\vert \ell ^{2})}%
\right].
\end{equation}%

At this point, we offer several important comments. First, the expression in Eq.~(\ref{proo}) reveals that the pair creation probability acquires an additional contribution due to the deformation parameter $\ell$, which originates from the spin noncommutativity of coordinates. This correction arises from modifications to the standard Heisenberg algebra involving spinor degrees of freedom. Notably, the exponent in Eq.~(\ref{proo}) vanishes when $\ell = \frac{m}{eE}$, leading to a formal divergence in the pair creation probability. Physically, this behavior identifies a critical value of the deformation parameter beyond which the vacuum becomes highly unstable. In this regime, even a relatively weak electric field may induce prolific pair production, indicating a possible breakdown of the semiclassical approximation or the emergence of strong backreaction effects.

Furthermore, even for small values of $\ell$, spin noncommutativity enhances the pair creation probability compared to the standard Schwinger result. This enhancement could have significant implications for strong-field QED, particularly in the context of future high-intensity laser experiments aimed at probing nonperturbative quantum vacuum phenomena.

The divergence of the pair creation probability at the critical value $\ell = \frac{m}{eE}$ carries important physical meaning. As Eq.~(\ref{proo}) shows, the exponential suppression factor in the Schwinger-like expression disappears at this threshold, leading to a divergent sum over pair production rates. Mathematically, this corresponds to a logarithmic divergence stemming from the harmonic series. Physically, it implies that the vacuum becomes maximally unstable when the deformation parameter reaches this critical value. In such a regime, the effective mass squared of the fermion field, as modified by spin noncommutativity, vanishes:
\begin{equation}
m_{\text{eff}}^2 = m^2 - e^2 E^2 \ell^2 \longrightarrow 0.
\end{equation}
This indicates that the external field, when combined with the deformation, becomes strong enough to close the mass gap, thereby enabling spontaneous and unsuppressed pair production. Such behavior may signal a breakdown of the semiclassical approximation employed in the proper-time formalism and could necessitate a fully quantum treatment of the vacuum. Alternatively, it may suggest that the deformation parameter $\ell$ must be restricted to remain below the critical threshold in any physically consistent theory. In this context, the divergence serves as a theoretical bound, delineating the limit beyond which the vacuum structure in spin noncommutative geometry becomes dynamically unstable.

\begin{figure}[H]
    \centering
    \begin{subfigure}{.35\textwidth}
        \centering
        \includegraphics[width=1.0\linewidth]{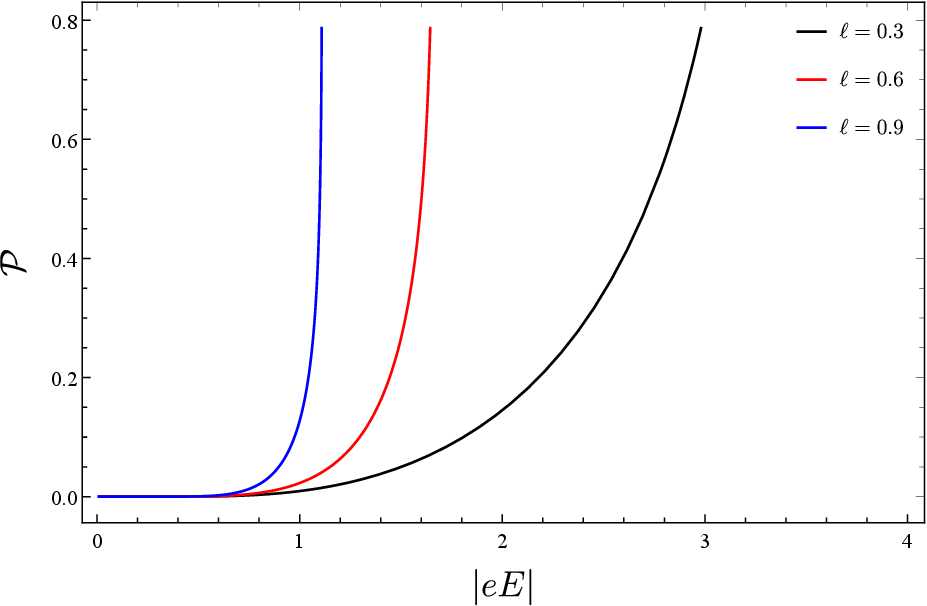}
        \caption{$\mathcal{P}$ vs. $\vert eE\vert$}
        \label{sub1}
    \end{subfigure}
    \hfill
    \begin{subfigure}{.35\textwidth}
        \centering
        \includegraphics[width=1.0\linewidth]{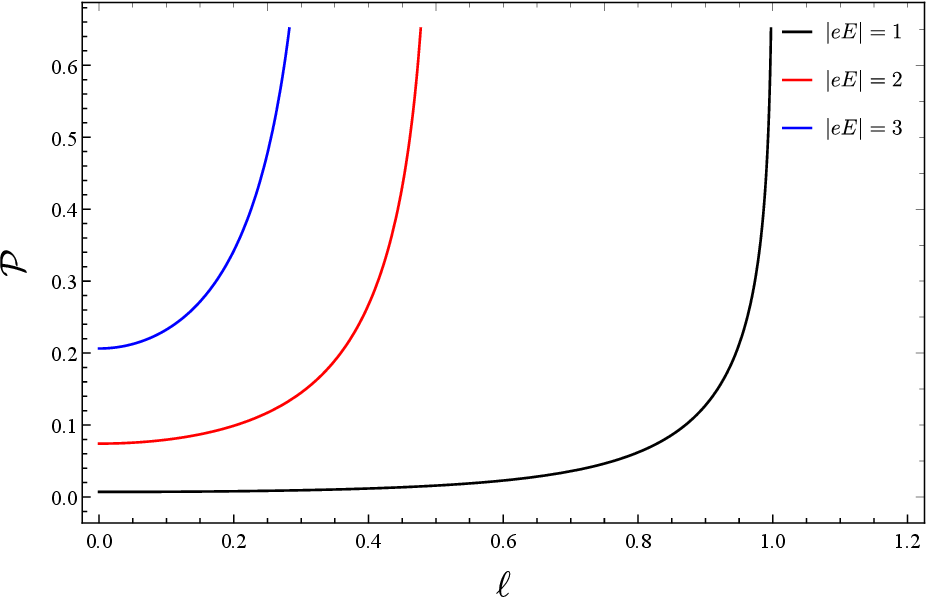}
        \caption{$\mathcal{P}$ vs. $\ell$}
        \label{sub2}
    \end{subfigure}
    \hfill
    \begin{subfigure}{.35\textwidth}
        \centering
        \includegraphics[width=1.0\linewidth]{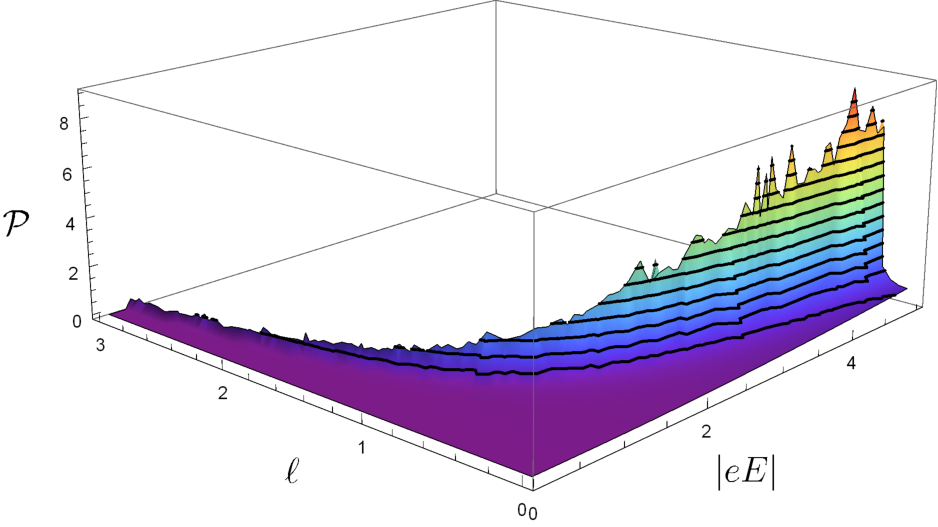}
        \caption{$\mathcal{P}$ vs. $\ell$ and $\vert eE\vert$}
        \label{sub3}
    \end{subfigure}
    \caption{Dependence of the pair creation probability $\mathcal{P}$ on the electric field strength $\vert eE\vert$ and the spin noncommutativity parameter $\ell$.}
    \label{fig:ueff}
\end{figure}

Figure~\ref{fig:ueff} illustrates the dependence of the pair creation probability $\mathcal{P}$ on the electric field strength $\vert eE\vert$ and the spin noncommutativity parameter $\ell$. 

In Figure~\ref{sub1}, for fixed values of $\ell$, the probability $\mathcal{P}$ increases with the electric field strength $\vert eE\vert$ and diverges at the critical value $\vert eE\vert = \frac{m}{\ell}$. This divergence corresponds to a vanishing effective mass gap, beyond which the vacuum becomes unstable. Figure~\ref{sub2} shows $\mathcal{P}$ as a function of $\ell$ for different values of $\vert eE\vert$, revealing a similar critical behavior: the probability grows monotonically with $\ell$ and diverges when $\vert eE\vert = \frac{m}{\ell}$. These findings confirm that spin noncommutativity enhances the pair creation process in a nonlinear fashion. Specifically, larger values of $\ell$ amplify the production rate for a given field strength, while stronger electric fields lower the threshold value of $\ell$ at which vacuum instability sets in. Figure~\ref{sub3} presents a three-dimensional surface plot of $\mathcal{P}$ as a function of both $\ell$ and $\vert eE \vert$. This visualization clearly captures the nonlinear interplay between the two parameters, highlighting how the pair creation probability increases rapidly as either variable approaches its respective critical threshold and emphasizing the sharp transition into the instability regime.

Finally, although the analysis presented here was conducted in $(1+1)$-dimensional spacetime for analytical simplicity, it would be highly valuable to extend this study to the physically relevant $(3+1)$-dimensional case. In such a setting, additional effects could arise from transverse momentum components, spin projections along multiple spatial directions, and the presence of magnetic fields in more general electromagnetic configurations. These factors would enrich the structure of the effective action and could significantly impact the pair production rate. While the proper-time formalism remains applicable, the algebraic complexity increases due to the full Dirac spinor structure in four dimensions. Nevertheless, such an extension would offer a more comprehensive understanding of how spin noncommutativity modifies nonperturbative QED processes in realistic environments, including high-intensity laser systems and extreme astrophysical conditions.

\section{Conclusion} \label{sec6}

In this work, we investigated the vacuum pair creation of Dirac particles in external electromagnetic backgrounds within the framework of spin noncommutativity of coordinates. Employing the Schwinger proper-time formalism, we computed the one-loop effective action and derived closed-form expressions for the pair creation probability in two distinct configurations: a Volkov plane wave and a constant electric field.

Our analysis confirms that, consistent with standard QED, no pair production occurs in a Volkov plane wave background—even when spin noncommutativity is introduced. This result reinforces the known nonperturbative stability of such backgrounds in both commutative and spin-noncommutative geometries.

In contrast, the presence of a constant electric field in $(1+1)$-dimensional spacetime leads to significant modifications once spin noncommutativity is taken into account. The deformation parameter $\ell$ appears explicitly in the exponent of the Schwinger-like pair production formula, enhancing the creation probability. Notably, we identified a critical value $\ell = \frac{m}{\vert eE \vert}$ at which the effective mass vanishes and the pair creation probability diverges—signaling a vacuum instability. This suggests that spin noncommutativity acts as an amplifier of the Schwinger effect, lowering the energy barrier for vacuum decay and potentially modifying the strong-field dynamics of the vacuum itself.

These results highlight the potential phenomenological relevance of spin noncommutativity in nonperturbative QED processes, particularly in light of ongoing and future high-intensity laser experiments where extreme field strengths may make such geometric effects observable.

Although our analysis was confined to a $(1+1)$-dimensional model to retain analytical tractability, a natural and important extension would be to generalize the study to $(3+1)$ dimensions. This would allow for the inclusion of transverse momenta, full spinor dynamics, and more general electromagnetic configurations, offering a deeper understanding of how spin-induced noncommutativity alters quantum vacuum phenomena in realistic physical settings.

\section*{Acknowledgments}
B. C. L. is grateful to Excellence Project PřF UHK 2205/2025-2026 for the financial support.

\section*{Data Availability Statements}

The authors declare that the data supporting the findings of this study are
available within the article.

\end{document}